# Small Drone Classification with Light CNN and New Micro-Doppler Signature Extraction Method Based on A-SPC Technique

Junhyeong Park, *Member, IEEE*, Jun-Sung Park, and Seong-Ook Park, *Senior Member, IEEE*

*Abstract*—As the threats of small drones increase, not only the detection but also the classification of small drones has become important. Many recent studies have applied an approach to utilize the micro-Doppler signature (MDS) for the small drone classification by using frequency modulated continuous wave (FMCW) radars. In this letter, we propose a novel method to extract the MDS images of the small drones with the FMCW radar. Moreover, we propose a light convolutional neural network (CNN) whose structure is straightforward, and the number of parameters is quite small for fast classification. The proposed method contributes to increasing the classification accuracy by improving the quality of MDS images. We classified the small drones with the MDS images extracted by the conventional method and the proposed method through the proposed CNN. The experimental results showed that the total classification accuracy was increased by 10.00 % due to the proposed method. The total classification accuracy was recorded at 97.14 % with the proposed MDS extraction method and the proposed light CNN.

*Index Terms*—Advanced stationary point concentration (A-SPC) technique, convolutional neural network (CNN), drone classification, frequency modulated continuous wave (FMCW) radar, leakage mitigation, micro-Doppler signature (MDS).

## I. Introduction

AS the drone market grows, drones are getting smaller and smarter. To defend against the threats from small drones, the identification as well as the detection of the small drones has become important. In recent papers, an approach to classify the small drones based on their micro-Doppler signature (MDS) has been often used [1]–[7]. Because the MDS is formed differently depending on the number, rotational frequency, and shape of propellers, the MDS has been used for the small drone classification. For this approach, frequency modulated continuous wave (FMCW) radars have been frequently used [3]–[7]. However, the FMCW radar has an inherent problem called leakage. The leakage signal generated by the leakage to the receiver of the transmitted signal significantly reduces the sensitivity of the FMCW radar [8]–[11].

To mitigate the leakage signal, we have proposed a new technique called stationary point concentration (SPC) technique [8]–[10], and recently we have proposed the advanced SPC (A-SPC) technique that overcomes the shortcomings of the SPC technique [11]. However, the SPC technique or the A-SPC technique has only been used for the detection of small drones. Although we have raised and shown the feasibility that these techniques can be used as basic techniques for various FMCW radar applications in addition to simple target detection [10], [11], none of the applications based on these techniques has been performed or proposed yet.

This letter is the first case to verify that the A-SPC technique can also be used for more advanced FMCW radar applications than the simple target detection. In this letter, we propose a novel method based on the A-SPC technique for the improvement of MDS images to classify the small drones. The proposed method makes the feature of MDS clear, so it can increase the total classification accuracy of the classifier that is a convolutional neural network (CNN) in this paper. A light CNN whose the number of parameters is considerably small is also proposed for the fast classification.

For the experiments, a *Ku*-band FMCW radar was used, and three different small drones were used as targets. We extracted the MDS images of the small drones and noises by both the conventional method and the proposed method. Then, we performed the classification of five classes that are three classes of the small drones, and two classes of the noises by using their MDS images. The classification results based on the MDS images extracted by the conventional method were compared with those based on the MDS images extracted by the proposed method.

We show that the MDS images of the small drones have been improved by the proposed method. Then, we verify that the total classification accuracy has been increased due to the MDS images improved by the proposed method. Finally, an interesting result is shown, which is that if the proposed method is used, fairly high classification accuracy can be achieved even with a quite simple CNN, such as the proposed light CNN.

## II. Proposed Method and Proposed Light CNN

### A. Proposed MDS extraction method

The proposed method is shown in Fig. 1 as a form of flow chart. In the FMCW radar, the phase noise of the leakage signal dominates the noise floor and reduces the signal-to-noise ratio (SNR) of targets [8]-[11]. The A-SPC technique included in the proposed method finds the leakage signal and concentrates its phase noise on the stationary point in the sinusoidal function of the leakage signal [11]. These procedures are shown in Fig. 1 as a part of the proposed method. Quadrature raw data, $x_{raw}[n,m]$, go through the process of the quadrature imbalance correction. Then, the index number for the leakage signal, $k_{IF\ leakage}(m)$, is found from the power spectrum of the corrected data, $x[n,m]$. Note that $X[k,m]$ is the result of the $NFFT$-point fast Fourier transform (FFT) of $x[n,m]$ along the fast time domain whose index is expressed as $n$. $NFFT$ is the total number of data samples and zero-pads for the zero-padding. Then, the frequency, $f_{IF\ beat\ leakage}(m)$ and the phase, $\theta_{IF\ leakage}(m)$, values of the leakage signal can be found as presented in Fig. 1. $F_S$ is the sampling frequency, and $\angle X$ is the phase response of $X[k,m]$. After generating a digital numerically controlled oscillator (NCO) with the found $f_{IF\ beat\ leakage}(m)$ and $\theta_{IF\ leakage}(m)$, the complex-based mixing is conducted by taking the conjugate to $NCO[n,m]$ and multiplying it with $x[n,m]$. Finally, the real data in the output of the mixing are extracted. These procedures make the phase noise of the leakage signal concentrated on the stationary point, and lead to the significant attenuation in the magnitude of its phase noise, thus the noise floor is decreased, and the SNR is increased.

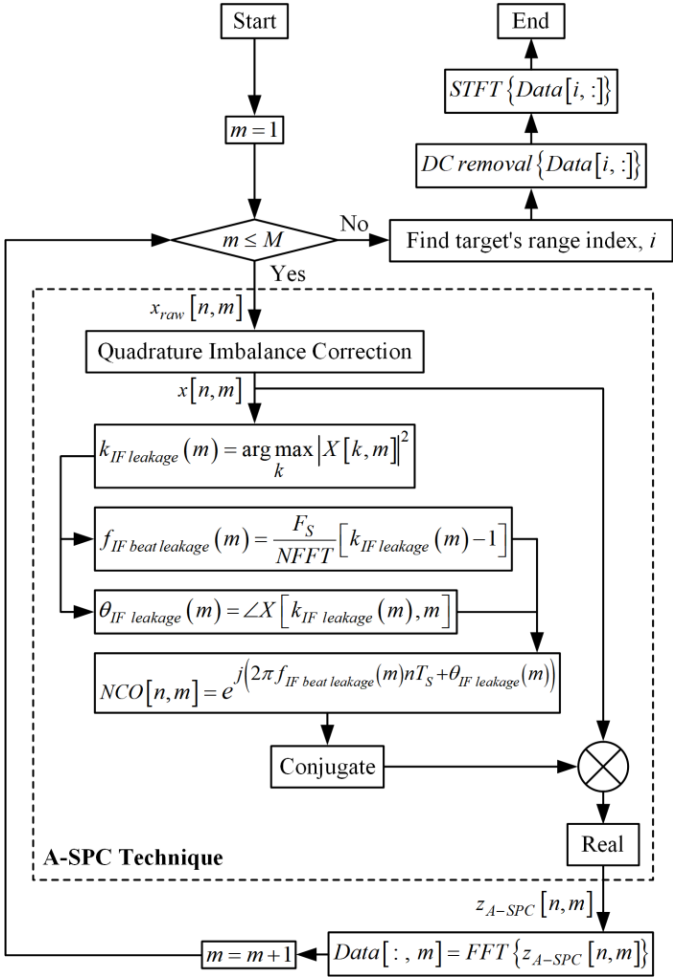

Fig. 1. Proposed MDS extraction method.

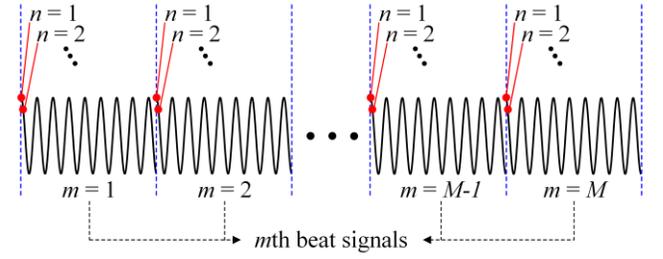

Fig. 2. Conceptual figure to explain the notations of $n$ and $m$.

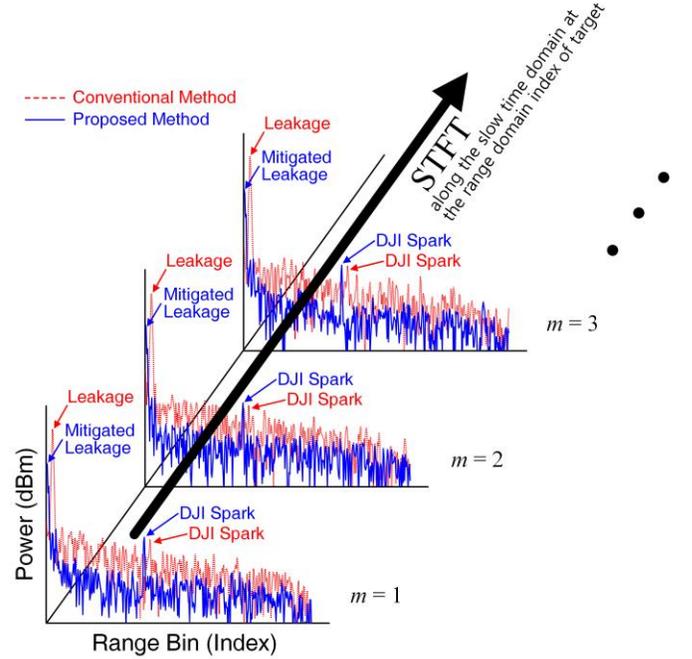

Fig. 3. Supplementary figure to explain the proposed method with some results in the middle of the process. The target was a hovering DJI Spark.

The proposed method continually repeats these procedures for every $m$th bunch of beat signals until $m$ reaches $M$, where $m$ is the index representing slow time domain, and $M$ is the total number of $m$ to extract an MDS image. Fig. 2 helps to explain the notations, $n$ and $m$. For each $m$th bunch of beat signals, the proposed method applies the FFT to the output of the aforementioned procedures, $z_{A-SPC}[n,m]$, then accumulates the FFT results. When this iteration is finished, the proposed method removes the dc values and applies short-time Fourier transform (STFT) along the slow time domain at the range domain index, $i$, of the target. Finally, the MDS image can be obtained by plotting the spectrogram that is the result of STFT.

Fig. 3 shows some results in the middle of the process. Since the proposed method repeats the procedures of the A-SPC technique for every $m$th bunch of beat signals, the leakage signal in each $m$th bunch of beat signals is mitigated. Thus, the SNR of the target in each $m$th bunch of beat signals is increased, and this naturally increases the SNR of the target along the slow time domain. Therefore, the MDS image can be improved. Note that the reason why the peak of the target is shifted in the spectrum for the proposed method in Fig. 3 is due to the additional function of the A-SPC technique, which compensates for the internal delay of the radar system [11].

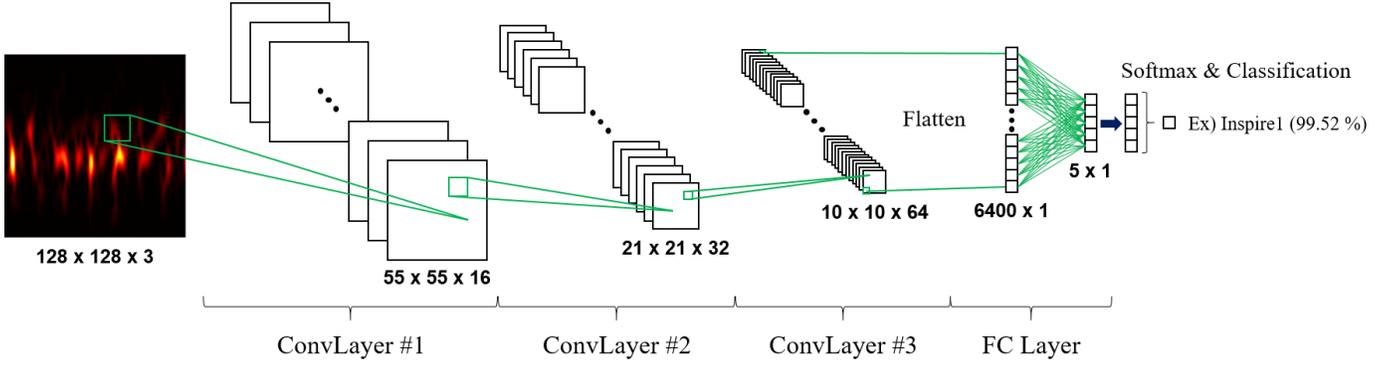

Fig. 4. Structure of the proposed light CNN.

TABLE I
DETAILED STRUCTURE OF THE PROPOSED LIGHT CNN

| | Parameters | | Values |
|---|---|---|---|
| ConvLayer #1 | Convolution | Filter size | 21 × 21 |
| | | Padding | 1 |
| | | Stride | 1 |
| | | Depth (# of bias) | 16 |
| | Activation | Function | ReLU |
| | Pooling | Type | Max pooling |
| | | Filter size | 2 × 2 |
| | | Stride | 2 |
| ConvLayer #2 | Convolution | Filter size | 16 × 16 |
| | | Padding | 1 |
| | | Stride | 1 |
| | | Depth (# of bias) | 32 |
| | Activation | Function | ReLU |
| | Pooling | Type | Max pooling |
| | | Filter size | 2 × 2 |
| | | Stride | 2 |
| ConvLayer #3 | Convolution | Filter size | 4 × 4 |
| | | Padding | 1 |
| | | Stride | 1 |
| | | Depth (# of bias) | 64 |
| | Activation | Function | ReLU |
| | Pooling | Type | Max pooling |
| | | Filter size | 2 × 2 |
| | | Stride | 2 |
| FC Layer | Filter size | | 6400 × 5 |
| | # of bias | | 5 |
| Classifier | | | Softmax |

*B. Proposed light CNN*

There are two approaches for the fast classification process with the CNN. The first is to increase the performance of processing units, such as CPU and GPU, and the second is to reduce the total number of parameters of the CNN so that the CNN becomes simple. Because the first one depends on the budget, the second approach is more reasonable and preferred. Thus, we propose a light CNN, as shown in Fig. 4. The proposed light CNN consists of only three convolutional layers and one fully-connected layer. The detailed structure of the light CNN is presented in Table I. For the activation in each convolutional layer, the Restricted Linear Units (ReLU) is used, and the Softmax is used as the classifier at the end of the CNN. The number of parameters for three convolutional layers is 185,120, and the number of parameters for a fully-connected layer is 32,005. Therefore, the total number of parameters for the proposed light CNN is 217,125.

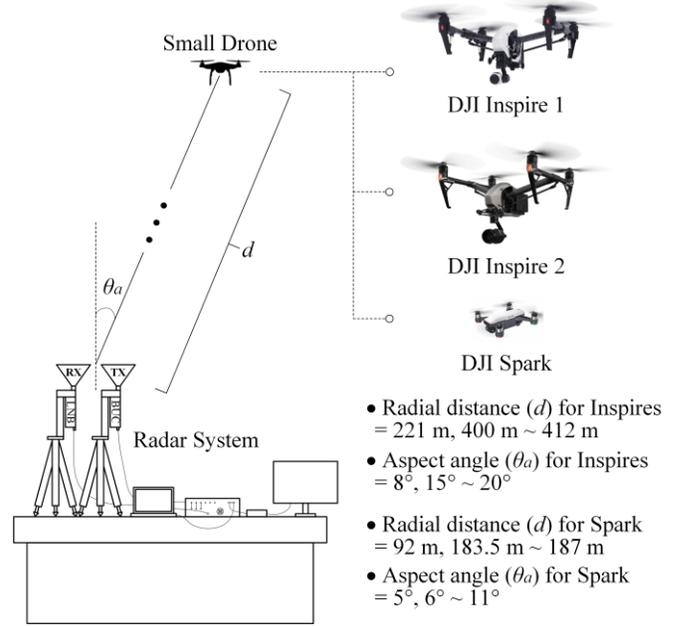

Fig. 5. Experimental setup and targeted small drones.

- Radial distance ($d$) for Inspires
  = 221 m, 400 m ~ 412 m
- Aspect angle ($\theta_a$) for Inspires
  = 8°, 15° ~ 20°
- Radial distance ($d$) for Spark
  = 92 m, 183.5 m ~ 187 m
- Aspect angle ($\theta_a$) for Spark
  = 5°, 6° ~ 11°

## III. EXPERIMENTS

Fig. 5 shows the experimental setup and targeted small drones. The experiments were conducted on the rooftop of a building. The *Ku*-band FMCW radar was used and commercial small drones, DJI Inspire 1, DJI Inspire 2, and DJI Spark, were used as targets. The detailed specifications and parameters of the FMCW radar used in the experiments are listed in Table II. We measured each small hovering drone at various radial distances and aspect angles as described in Fig. 5. Then, by using the conventional method and the proposed method, respectively, we extracted the colored MDS images whose size is 128×128×3. We tried to classify a total of five classes that are three classes of small drones and two classes of noises. One noise class, Noise 1, is the MDS image of the noise when the parameters for the STFT are set to those for DJI Inspires as listed in Table II. The other noise class, Noise 2, is the MDS image of the noise when the parameters for the STFT are set to those for DJI Spark as listed in Table II. Namely, we tried to check the classification performance not only between the small drones but also between the targets and noise. We collected 700 MDS images per class, resulting in a total of 3,500 MDS images. Among the 3,500 MDS images, we used 80 % of them for training and 20 % of them for validation. For the training of the light CNNs, the Adam optimizer with the learning rate of 0.0001 was used. The batch size was 70, and the number of epochs was 100. Finally, we tested the trained CNNs with the newly extracted 1,750 MDS images that are independent of the training data.

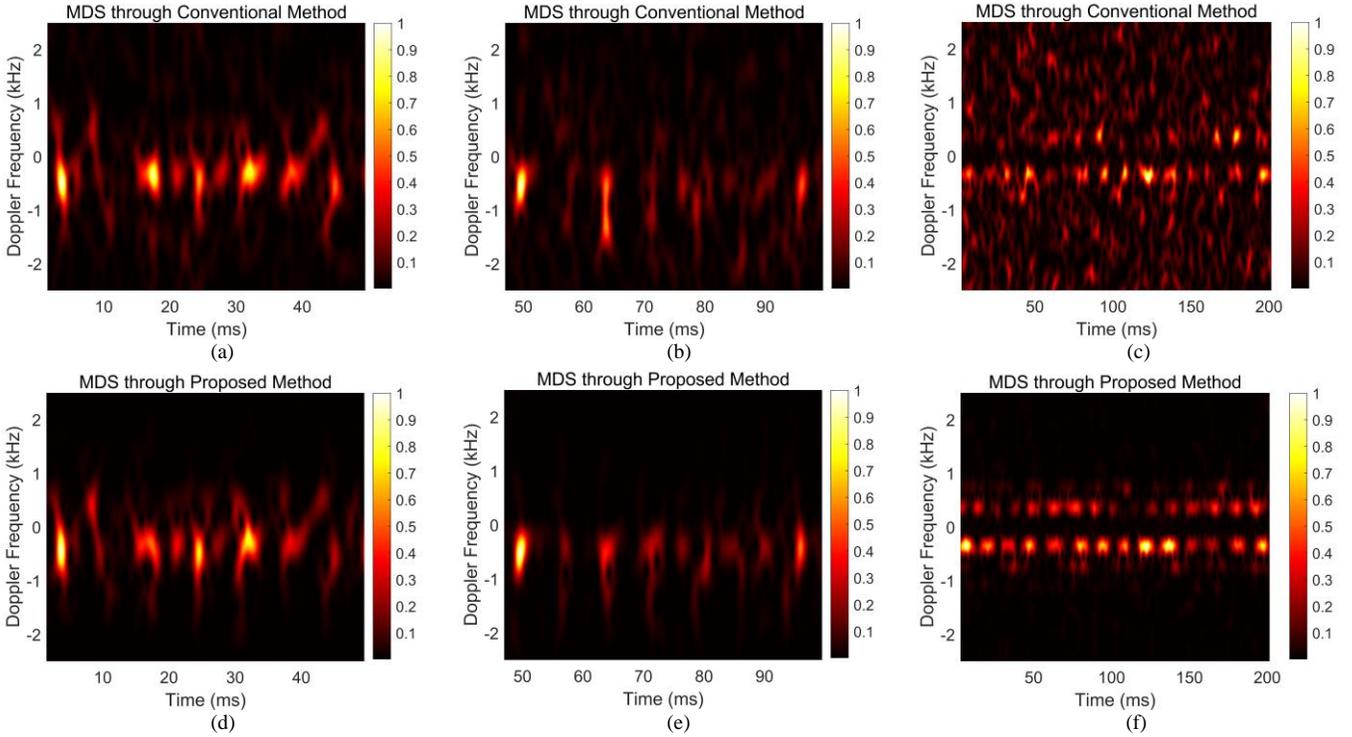

Fig. 6. Extracted representative MDS images. (a)-(c) MDS images through the conventional method. (d)-(e) MDS images through the proposed method. (a) and (d) are the MDS images of DJI Inspire 1. (b) and (e) are the MDS images of DJI Inspire 2. (c) and (f) are the MDS images of DJI Spark.

TABLE II
SPECIFICATIONS AND PARAMETERS OF THE FMCW RADAR

| Parameters | Values |
|---|---|
| Radar configuration | Quasi-monostatic |
| System architecture | Heterodyne |
| Operating frequency | 14.35-14.50 GHz |
| Transmit power | 30 dBm |
| Antenna gain | 16 dBi |
| Sweep bandwidth ($BW$) | 150 MHz |
| True range resolution | 1 m |
| Final IF carrier frequency ($f_{IF\ carrier}$) | 0 MHz |
| Sweep period ($T$) | 200 us |
| Desired digital bandwidth | 2.5 MHz |
| Sampling frequency ($F_S$) | 5 MHz |
| Samples in a chirp | 1000 |
| Desired maximum detectable range | 500 m |
| NFFT for finding out $f_{IF\ beat\ leakage}$ and $\theta_{IF\ leakage}$ | $2^{19}$ |
| Window for fast time domain | Hann |
| Chirps for MDS extraction of DJI Inspires ($M$) | 256 |
| Chirps for MDS extraction of DJI Spark ($M$) | 1024 |
| Window for slow time domain | Hann |
| Window length of STFT for DJI Inspires | 16 |
| Window length of STFT for DJI Spark | 32 |
| Overlapped sample in sliding window for DJI Inspires | 15 |
| Overlapped sample in sliding window for DJI Spark | 31 |
| Renewed sample in sliding window | 1 |

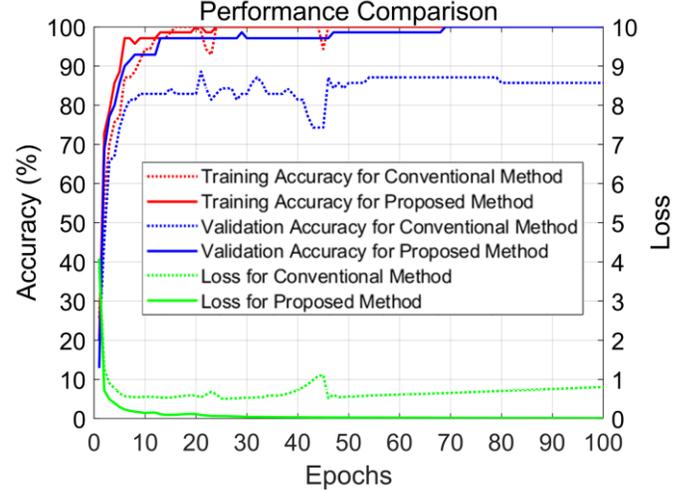

Fig. 7. Comparison of the training performance.

## IV. RESULTS AND DISCUSSION

Fig. 6 shows representative MDS images extracted by each method. As we expected, the proposed method successfully improves the MDS images by reducing noises in the image and making the feature of the MDS clear. Fig. 7 shows the training performance when we train the light CNN with the MDS image data extracted by each method. As shown in Fig. 7, the training accuracy based on the data extracted by the proposed method rises faster than that by the conventional method. In addition, the validation accuracy and the loss due to the proposed method are clearly better than the conventional method.

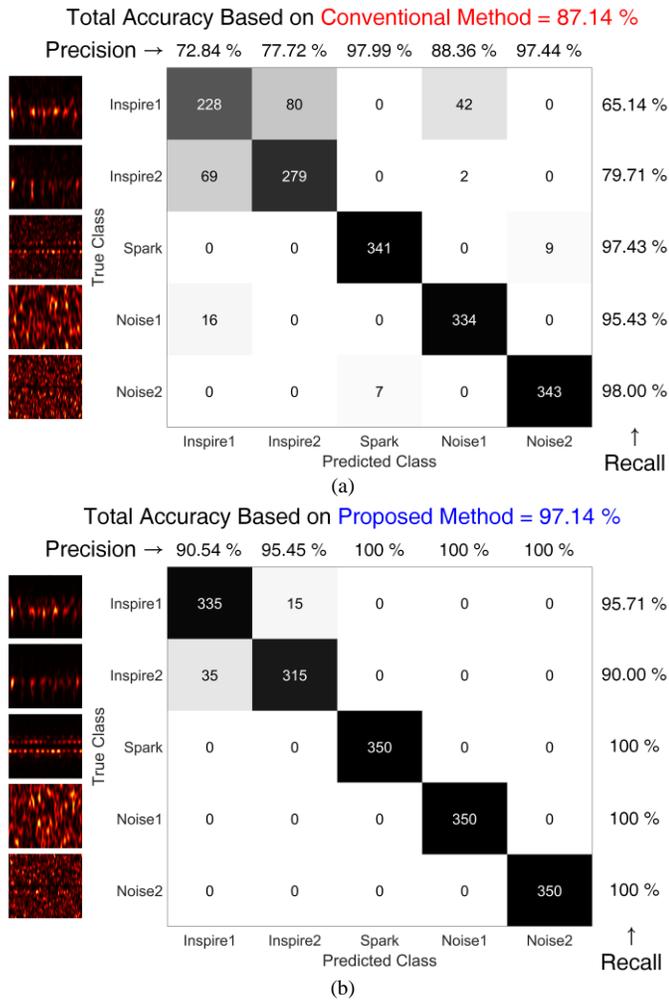

Fig. 8. Final classification results. (a) Confusion map based on the results from the conventional method. (b) Confusion map based on the results from the proposed method.

Fig. 8 shows the final classification results. In the case of the results based on the conventional method as shown in Fig. 8(a), the classification performance of DJI Spark and noise classes is not bad. However, the classification performance for DJI Inspire 1 and DJI Inspire 2 is poor. The trained CNN has not classified DJI Inspire 1 and DJI Inspire 2, which are very similar to each other, well. Besides, even the classification between the target and the noise is not good in the case of DJI Inspire 1. The total classification accuracy is 87.14 %. On the other hand, in the case of the results based on the proposed method as shown in Fig. 8(b), the overall classification performance is great. The trained CNN perfectly classifies DJI Spark and noise classes, and there is no case of failure to classify between the target and the noise. Moreover, the classification performance for DJI Inspire 1 and DJI Inspire 2, which are hard to classify due to their similar shape, is much better than the results of the conventional method. The total classification accuracy is 97.14 %. Thus, the proposed method can improve the total accuracy by 10 %. As listed in Table III, we compared our performances with the results included in the most popular paper that has the same research topic. Even though the number of the parameters of the proposed light CNN is much smaller than that of the CNN in [3], the total classification accuracy is higher.

TABLE III
COMPARISON BETWEEN THE MOST POPULAR PAPER AND THIS WORK

| Paper | Data Type | MDS Extraction Method | CNN | # of CNN Parameters | Total Accuracy |
|---|---|---|---|---|---|
| [3] | MDS | Conventional Method | GoogLeNet | 6.8 M | 89.3 % |
| This work | MDS | Proposed Method | Proposed Light CNN | 217 K | 97.14 % |

V. CONCLUSION

We have proposed the MDS extraction method and the light CNN for the small drone classification. Experimental results have proved that the proposed MDS extraction method improves the quality of the MDS image, thus it increases the total classification accuracy. In addition, high classification accuracy has been recorded with the proposed light CNN whose the number of parameters is considerably small, when the proposed method is applied together. Therefore, it has been verified that the combination of the proposed method and the proposed light CNN is a quite useful solution for the fast and accurate classification of the small drones in practice.